\documentclass[preprint,prb,aps,showpacs,preprintnumbers,amsmath,amssymb,superscriptaddress,floatfix,footinbib]{revtex4}
\usepackage{graphicx}
\usepackage{dcolumn}
\usepackage{bm}
\begin{document}
\title{Hartree-Fock theory of a current-carrying electron gas}
\author{H. Mera}
\affiliation{Department of Physics, University of York, Heslington, York YO10
5DD, United Kingdom.}
\affiliation{Niels Bohr Institute and Nano-Science Center, Universitetsparken 5,
DK-2100 Copenhagen Ø.}
\author{P. Bokes}
\affiliation{Department of Physics, University of York, Heslington, York YO10
5DD, United Kingdom.}
\affiliation{Department of Physics, Faculty of Electrical Engineering and Information Technology, Slovak University of Technology,
Ilkovi\v{c}ova 3, 812 19 Bratislava, Slovakia.}
\author{R.W. Godby}
\affiliation{Department of Physics, University of York, Heslington, York YO10
5DD, United Kingdom.}
\date{\today}
\begin{abstract}
State-of-the-art simulation tools for non-equilibrium quantum transport systems typically take the
	current-carrier occupations to be described in terms of 
	equilibrium distribution functions
	characterised by two different electro-chemical potentials, while for the description
	of electronic exchange and correlation, the local density approximation (LDA) to density functional 
	theory (DFT) is generally used.
	However this involves an inconsistency because the LDA is based on the homogeneous electron gas \emph{in
	equilibrium}, while the system is not in equilibrium and may be far from it.
	In this paper we analyze this inconsistency by studying the interplay between non-equilibrium
	occupancies obtained from a maximum entropy approach and the Hartree-Fock
	exchange energy, single-particle spectrum and exchange hole, for the case of a two-dimensional homogeneous
	electron gas. The current-dependence of the local exchange potential is also discussed. It is found that
	the single-particle spectrum and exchange hole have a significant dependence on the 
	current which has not been taken into account in
	practical calculations. The exchange energy and the 
	local exchange potential, however, are shown to change very little with respect to their equilibrium counterparts.
	The weak dependence of these quantities on the current is explained in terms of the symmetries of the
	exchange hole.
\end{abstract}
\maketitle
\section{Introduction}
	One of the uncontrollable approximations introduced in {\it ab initio} calculations of the transport
	properties of nano-scale conductors consists in the application of DFT, a ground state theory, outside the
	equilibrium regime. An immediate consequence of this approximation is that these properties are
	 typically calculated
	at the level of the LDA, which is derived from the case of
	a homogeneous electron gas in equilibrium. The extent to which these approximations might affect the
	calculated electronic structure of the non-equilibrium systems remains largely unknown and thus
	a comparison between electronic properties calculated \emph{exactly} for an admittedly highly idealised
	non-equilibrium system and those of the same system in equilibrium 
	constitutes a particularly simple way of approaching and illustrating this problem.
	
	In order to put these ideas into practice we will consider a two-dimensional electron gas in equilibrium
	and in a \emph{model} non-equilibrium state. To model a homogeneous
	electron gas outside equilibrium we will assume that the 
	non-equilibrium steady-state of the two dimensional electron gas can be characterized by the average 
	total energy
	of the electron gas and by \emph{different} average numbers of left- and right-moving electrons and 
	that the non-equilibrium steady-state is given by the density matrix that maximises the entropy
	of the electron gas with constraints on the above mentioned averages. 
	
	Such an assumption leads in the
	non-interacting case to a momentum distribution characterized by \emph{two Fermi hemispheres of different
	radii};
	we take a pragmatic approach here and ignore the problems associated with the discontinuous character of this momentum
	distribution for the time being since we are 
	interested in the question of how these current-inducing constraints affect the electronic properties of
	the two dimensional electron gas. Note that this type of momentum distribution is
	precisely of the form used in Landauer-B\"uttiker-type of approaches and thus familiar to the 
	{\it ab initio} quantum transport community \cite{dattabook,JauhoBook,Taylor1,Taylor2} which constantly makes use of it. 
	Similar momentum distributions are predicted by semi-classical transport theories in two dimensional quantum point
	contacts \cite{Jansen1980}. Alternatively, and perhaps also more physically, 
	a current-constraint may be used instead of the
	above-mentioned constraint to search for the non-equilibrium maximum entropy density matrix \cite{Ng92,Heinonen93,Bokes03,Bokes05a}.

	To summarise, we will maximize the entropy of a two-dimensional 
	homogeneous electron gas with constraints on the 
	average 
	 numbers of left- and right-moving electrons to obtain a description of a steady-state at the
	 Hartree-Fock level of approximation, which can then be used to obtain the electronic structure of the gas
	 in the presence of a current and to compare it with the usual approximations. The rest of the paper is organized as follows: in
	 the next section we discuss our theoretical approach to the problem and its numerical implementation; in Section III we discuss the
	 current-dependence of the Hartree-Fock pair probability distribution, single particle spectrum, total energy and local 
	 exchange potential. We conclude with a discussion of the relevance of our work for practical calculations.
	 
	 \section{Theory}

	In order to proceed let us consider the entropy per unit area of the two dimensional electron gas to be a functional of the electronic
	occupancies given by \cite{Todorov00}:
	\begin{equation}
	\label{eq:entropy}
	S\left[ f\right(\mathbf{k}\left)\right]
	=-
	\int_{\Re^2} \frac{d^2\mathbf{k}}{2\pi^2} \, \left[
	f(\mathbf{k})\ln f(\mathbf{k})+\left(1-f(\mathbf{k}) \right) \ln \left( 1-f(\mathbf{k})
	\right)
	\right] .
	\end{equation}
	The electronic occupancies are written as:
	\begin{displaymath}
        f(\mathbf{k})= \left\{\begin{array}{ll}
        f_L(\mathbf{k}) &    \textrm{if $k_x < 0$} \\
        f_R(\mathbf{k}) & \textrm{if $k_x > 0$}
        \end{array}\right.
	\end{displaymath}
	where $\mathbf{k}=\left(k_x , k_y \right)$ and $f_{L/R}$ are the occupation functions to be varied independently in order to 
	 maximise Eq.~(\ref{eq:entropy}) with
	constraints on the average total energy per unit area and different average numbers of left- and right-going particles per unit area.
	In the finite-temperature Hartree-Fock approximation the average total energy is given by:
	\begin{equation}
	\label{eq:hfetot}
	\langle E \rangle=2\int_{\Re^2}\frac{d^2\mathbf{k}}{\left( 2\pi \right)^2}\,f(\mathbf{k})\frac{k^2}{2}
	-\int_{\Re^2}\frac{d^2\mathbf{k}^\prime}{\left( 2\pi \right)^2}\int_{\Re^2}\frac{d^2\mathbf{k}}{\left( 2\pi \right)^2}
	f(\mathbf{k})f(\mathbf{k}^\prime)v(\mathbf{k},\mathbf{k}^\prime)
	\end{equation}	
where $v(\mathbf{k},\mathbf{k}^\prime)=2\pi/\left|\mathbf{k}-\mathbf{k}^\prime \right|$ is the Fourier transform of the Coulomb interaction
in two dimensions. The number of left- and right-going electrons per unit area can be written as:
\begin{equation}
\label{eq:densdes}
	n_{L(R)}=\frac{2}{\left( 2\pi\right)^2}\int_{k_x<(>)0} d^2\mathbf{k} \, f_{L(R)}(\mathbf{k})
\end{equation}
In order to maximise the entropy functional with respect to $f_{L/R}$ subject to the above-mentioned constraints we use the method of
Lagrange multipliers and consider the auxiliary functional
\begin{equation}
\mathcal{L}\left[ f\right(\mathbf{k}\left)\right]=S-\beta \left(
\langle E \rangle-\mu_L n_L -\mu_R n_R
\right)\, ,
\end{equation}
together with the extremal condition
\begin{equation}
\frac{\delta \mathcal{L}}{\delta f_{L/R}}=0 .
\end{equation}
A straightforward calculation shows that the occupation functions that maximise the entropy functional with constraints in the
above-mentioned averages are given by:
\begin{equation}
f_{L,R}(\mathbf{k})=\frac{1}{1+\exp \left[ \beta \left( 
k^2/2+\epsilon_x(\mathbf{k})-\mu_{L,R}
\right) \right]}
\end{equation}
where
\begin{equation}
\epsilon_x(\mathbf{k})= -\frac{1}{(2\pi)^2}\left(\int_{k_x<0} d^2\mathbf{k}^\prime \, f_L(\mathbf{k}^\prime) v(\mathbf{k},\mathbf{k}^\prime)
+\int_{k_x>0} d^2\mathbf{k}^\prime \, f_R(\mathbf{k}^\prime) v(\mathbf{k},\mathbf{k}^\prime)
\right)
\label{eq:xspec}
\end{equation}
i.e., the occupations that maximise the entropy are similar to the ones of the Landauer-B\"uttiker 
approach but with a modified exchange part
of the spectrum. 
In the calculation we fix the ratio $n_L/n_R$, that together with the charge neutrality condition $n_L+n_R=1/(\pi r_s^2)$ completely
determines both $n_L$ and $n_R$. With the equilibrium spectrum as a trial $\epsilon_x(\mathbf{k})$ we 
solve Eqs.~(\ref{eq:densdes}) for $\mu_L$ and $\mu_R$. With these
values of $\mu_{L,R}$ a new spectrum is constructed using Eq.~(\ref{eq:xspec}) and the iteration is completed and subsequently repeated until the input and output
spectra are identical to each other within the desired tolerance. All the results presented here are obtained in the $\beta \rightarrow
\infty$-limit, where our approach is equivalent to that of Hershfield \cite{Hershfield93} in the Hartree-Fock approximation \footnote{i.e., to find the
Slater determinant that minimizes the expectation value of the effective Hamiltonian $\widehat{F}=\widehat{H}_{HF}-\mu_L \widehat{N}_R-\mu_L
\widehat{N}_R$.}.
Once the self-consistent spectrum and occupation factors are obtained, other quantities like the exchange-energy and exchange hole can be
easily obtained. From these we can study how the local exchange potential of the electron gas depends on the current density
\footnote{
For two
Fermi-hemispheres of radii $k_L$ and $k_R$ the  
non-interacting electronic current is related to these densities by the expression $j=\frac{2}{3\sqrt{2\pi}}(n_L+n_R)(n_R^3-n_L^3)$}. 

\section{Results}
\subsection{Hartree-Fock pair distribution function}
\begin{figure*}[t]
\centering
\includegraphics[scale=1.25]{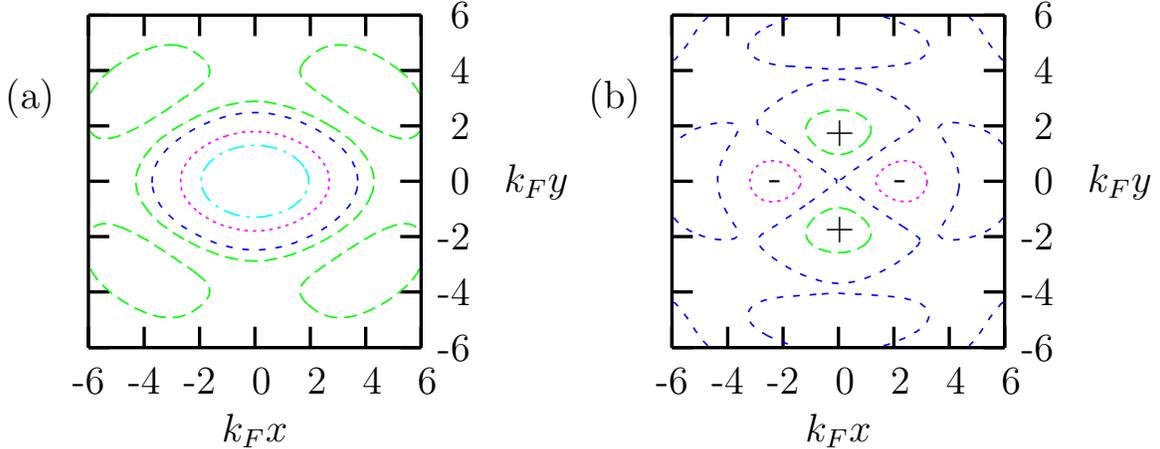}
\caption{(Color online). (a) Pair distribution function for like spins in the non-equilibrium regime ($n_L/n_R=0.5$). In the non-equilibrium
regime the exchange-hole is elongated along the direction of the current. The contours are at $g=$0.5, 0.75, 0.9
and 0.95. (b) Difference between the equilibrium and non-equilibrium holes, $\Delta g$ (see text). The contours are drawn at 0.1 (dashed), -0.1 (dotted) and
0 (dotted). $\Delta g$ is oscillating, integrates to zero and
has a marked antisymmetric character. Thus the current dependence of the local exchange potential and exchange energy is expected to be weak.}
\label{fig:hole}
\end{figure*}

Let us begin by discussing the current dependence of the Hartree-Fock pair distribution function for spin-like electrons, 
which is given by:
\begin{equation}
g(\mathbf{r},\mathbf{r^\prime})=1-\left| \frac{1}{n}\int \frac{d^2\mathbf{k}}{(2\pi)^2} 
\exp\left[-i\mathbf{k}\cdot (\mathbf{r-r^\prime})\right]f(\mathbf{k})\right|^2
\end{equation}
and shown for $n_L/n_R=0.5$ in Figure \ref{fig:hole}-(a). For $n_L=n_R$, $g$ is spherically symmetric
while for $n_L \neq n_R$ is elongated in the direction of the current. Similar phenomenology has been reported previously by
Skudlarski and Vignale for the three dimensional electron gas in the presence of a magnetic field \cite{sk93}, where the exchange hole is
elongated in the direction of the  field. In Ref.~\onlinecite{sk93} the elongation arises
from the change of occupancies associated with the Zeeman splitting due to the externally applied magnetic field.
In the present case the elongation of the hole can be understood in terms of the change in the electronic
occupancies that result from our constrained maximization of the entropy functional. In both cases the elongation of the hole is the result
of the change in the polarizability induced by the change in the occupancies \cite{sk93}.

Note that the difference
between the equilibrium and non-equilibrium exchange holes, $\Delta g=g_{eq}-g_{neq}$, shown in Figure \ref{fig:hole}-(b),
 has a strong antisymmetric character, i.e., defining $\mathbf{R}=\mathbf{r}-\mathbf{r}^\prime=(X,Y)$ then $\Delta g (X,Y)\sim -\Delta g
 (Y,X)$. We shall return to this point later in the text when discussing the weak dependence of the exchange energy on the current density.
\subsection{Single-Particle spectrum}

\begin{figure*}[t]
\includegraphics[scale=1]{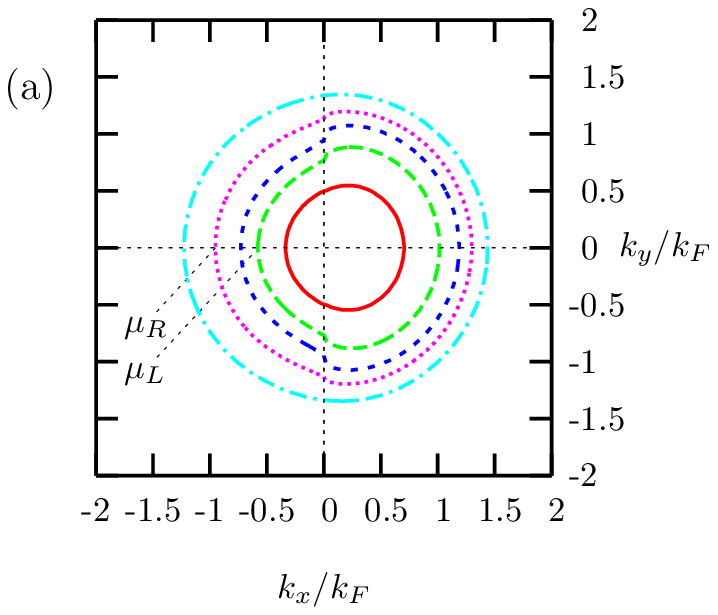}
\includegraphics[scale=0.95]{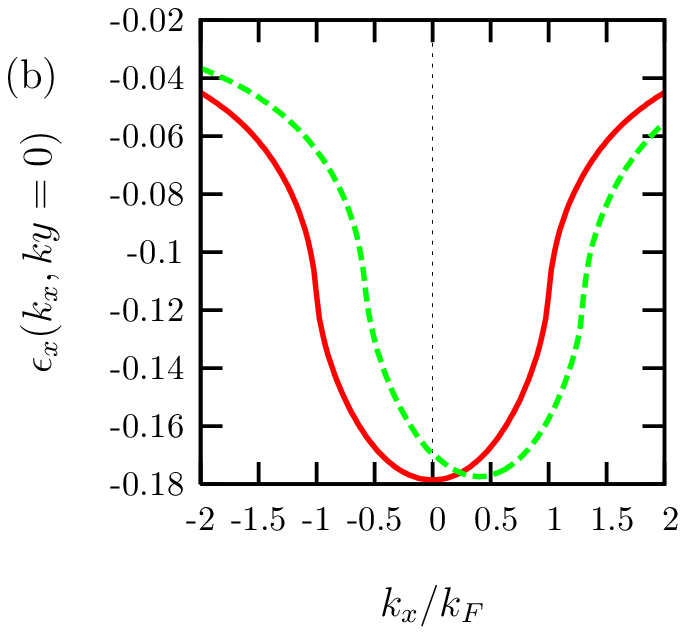} 
\caption{(Color online). (a) Contour plots of the total single particle energy spectrum of the model non-equilibrium electron gas for
$n_L/n_R=0.25$ and $r_s=4$. The contours corresponding to $\mu_L=-1.1\,10^{-2}$ a.u. and to $-9.2\,10^{-2}$ a.u. are labelled. 
The other contours shown correspond to
$(\mu_L+\mu_R)/2$ (short dashes), $\mu_L-0.5\,10^{-2}$(a.u.) (solid), $\mu_R+0.5\,10^{-2}$(a.u.) (dot-dashed). (b) 
Exchange contribution to the single particle energy spectrum, $\epsilon_x(\mathbf{k})$, evaluated on the $k_y=0$ line calculated for  $n_L=n_R$ (solid) and $n_L/n_R=0.25$ (dashed). The main effect 
of the non-equilibrium constraints used in our variational approach is to
	shift the exchange part of the single-particle spectrum towards higher values of $k_x$.}
\label{fig:scfspec}
\end{figure*}

	Figure~\ref{fig:scfspec} shows the self-consistent single-particle energy spectrum. Fig.\ref{fig:scfspec}-(a) shows the total
	(kinetic + exchange) spectrum while in Fig.\ref{fig:scfspec}-(b) we plotted only its exchange part 
	on the $k_y=0$ line  as given by Eq.~(\ref{eq:xspec})
	 for $n_L=n_R$ and $n_L/n_R=0.5$. 
	
	The combined effect of the constraints and the exchange interaction 
	shifts the spectrum towards 
	higher values of $k_x$. Note also that, when compared to the equilibrium spectrum, the minimum of the
	non-equilibrium spectrum is less negative. Hence we expect the total non-equilibrium 
	exchange energy to increase with respect to the equilibrium one. Note that the constraints alter 
	 the total kinetic energy of the system but do not change the kinetic
	 contribution to the single particle spectrum, since this contribution does not depend on the electronic
	 occupancies. Hence the changes in the single particle spectrum are entirely due to the exchange
	 interaction, which raises (lowers) the single particle energy of electrons with $k_x<0$  ($k_x>0$). The anomalous
	 behaviour in the $k_x=0$ plane inherited from the discontinuous character of the maximum entropy momentum distribution can be seen
	 clearly in Fig.\ref{fig:scfspec}-(a), between $\mu_L$ and $\mu_R$.

	The interplay between non-equilibrium occupancies and the single-particle 
	spectrum observed here is just a consequence of the orbital dependence of the Fock operator and 
	 will also be seen in any practical calculation that combines a non-equilibrium
	theory such as the Landauer-B\"uttiker approach or the Keldysh-NEGF formalism, with an \emph{orbital-dependent}
	description of the interactions between the electrons, such as the Hartree-Fock approximation. 
	We would like to point out that practical implementations of NEGF formalism typically take the
	electronic structure of the leads to be that of the equilibrium system (see Ref.\onlinecite{Mera05} and references
	therein), and hence the dependence of the single-particle spectrum on the non-equilibrium current (and
	vice-versa) is commonly ignored. The validity of this approximation is geometry dependent: it works in quantum point contact
	geometries while it does not in planar electrode geometries at high currents. As a consequence under the
	``non-interacting equilibrium lead approximation'' the distribution of incoming
	electrons would be current-independent, while, as this example shows\footnote{we can see our
	two-dimensional electron gas as a rough model of one of the leads to which the nanoscale conductor is
	attached}, the unavoidable presence of interactions in the leads induces a current dependence in the
	non-equilibrium occupancies through the exchange part of the single particle spectrum. Unless the geometry is adequately chosen
	the distribution of incoming electrons will be that of a \emph{non-equilibrium} lead such as ours.

\subsection{Total energy}

\begin{figure}[here]
\centering
\includegraphics[scale=1]{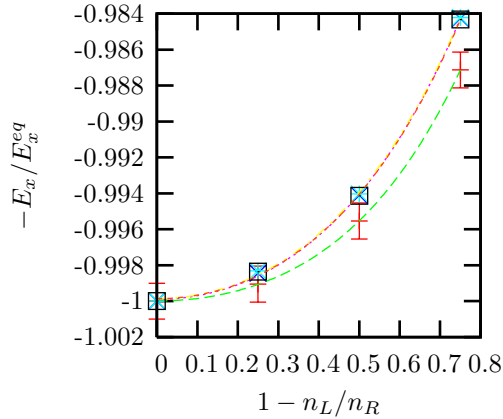}
\caption{(Color online). Exchange energy (in units of the
	equilibrium exchange energy) versus $1-n_L/n_R$. In equilibrium $1-n_L/n_R=0$. The dashed line shows
	the self-consistent results with estimated error bars. The non-self-consistent results are also shown with points calculated for
	different values of $r_s$ showing that the exchange energy scales with $r_s$ as $1/r_s$. The lines are fits to parabolic functions.}
	\label{fig:exvsj}
\end{figure}

	Once the self-consistent 
	single-particle spectrum is calculated the total exchange energy, $E_x$, 
	can be obtained from the second term in the right hand side of 
	Eq.~(\ref{eq:hfetot}). Figure \ref{fig:exvsj} shows the dependence of the $r_s$-invariant 
	quantity $-E_x/E_x^{eq}$ on $(1-n_L/n_R)$. 
	For $n_L/n_R=0.25$, the exchange energy deviates by about  $1-2\%$ from its equilibrium value. We also
	see that, even though the non-self-consistent results provide a good estimate to the self-consistent ones,
	full self-consistency is needed in the non-equilibrium case, even for a homogeneous gas. The error bars in
	the self-consistent results are estimated by comparing the exact exchange energy in equilibrium with the exchange energy obtained
	from our code for $n_L=n_R$ and different values of $r_s$.
	Therefore, the exchange-energy depends on the current-density, but this dependence is extremely weak in our
	model system.	One could now proceed to calculate this
	current density explicitly and work out a current dependent local density approximation, from the
	dependence of $E_x$ on the current density. However, the weak dependence of the exchange energy on the
	current density deduced from Fig. \ref{fig:exvsj} means that the current dependence of the local
	exchange functional is also very weak, and the changes it will induce in the associated LDA-Kohn-Sham
	effective potential will be well within the error bar of the LDA itself. 
	
\subsection{Local Exchange Potential}

	The weak dependence of the local exchange potential on the current density can be seen clearly in terms of
	the symmetries of the exchange hole. Consider the expression for Slater's exchange potential, $v^s_x$, in 
	terms of the Hartree-Fock
	pair distribution function:
	\begin{equation}
	v^s_x(\mathbf{r})=\int d^2 \mathbf{r'} \frac{[g(\mathbf{r}-\mathbf{r'})-1]}{|\mathbf{r}-\mathbf{r'}|}
	n(\mathbf{r'})
	\end{equation}
	where $n(\mathbf{r'})$ is the electron density and $g(\mathbf{r}-\mathbf{r'})$ is the exchange hole.
	Then, the difference between equilibrium and non-equilibrium exchange potentials is, for our homogeneous
	system, given by:
	\begin{equation}
	\Delta v^s_x =n \int d^2 \mathbf{R} \frac{\Delta g(\mathbf{R})}{|\mathbf{R}|}
	\label{eq:vxandhole}
	\end{equation}
	where $\mathbf{R}$ 
	and  $\Delta g$ are defined as above.
	From Eq.~(\ref{eq:vxandhole})  follows that:
	\begin{equation}
	\Delta g(X,Y)=-\Delta g(Y,X) \Rightarrow  \Delta v^s_x=0
	\end{equation}
	and hence only the symmetric part of $\Delta g(X,Y)$ contributes to the deviation of
	exchange potential with respect to its equilibrium value. Note that $\Delta g(X,Y)$ is an oscillatory function that integrates to zero 
	which also has a marked antisymmetric character shown in Fig.~\ref{fig:hole}-(b). This explains the weak dependence of 
	$E_x$ and $v_x$ on the current-density. 
\section{Conclusions}	
	
	In conclusion we have maximised the entropy of a two-dimensional homogeneous electron gas with constraints on the average 
	total energy and average numbers of left- and right-going electrons to obtain a simplified description of the steady-state within the
	Hartree-Fock approximation. Our results show that both the single-particle spectrum and the exchange hole depend significantly on the
	current density while averaged quantities like the local exchange potential or the exchange energy do not. 
	
\begin{acknowledgments}
The authors gratefully acknowledge useful discussions with
J.J. Palacios and J.Fern\'andez-Rossier. We are grateful to Matthieu Verstraete
for useful comments on the manuscript.
This work was supported by the EU's 6th Framework Programme 
through the NANOQUANTA Network of Excellence (NMP4-CT-2004-500198),
ERG programme of the European Union QuaTraFo (contract MERG-CT-2004-510615),
the Slovak grant agency VEGA (project No. 1/2020/05) 
and the NATO Security Through Science Programme (EAP.RIG.981521).
\end{acknowledgments}


\begin{thebibliography}{13}
\expandafter\ifx\csname natexlab\endcsname\relax\def\natexlab#1{#1}\fi
\expandafter\ifx\csname bibnamefont\endcsname\relax
  \def\bibnamefont#1{#1}\fi
\expandafter\ifx\csname bibfnamefont\endcsname\relax
  \def\bibfnamefont#1{#1}\fi
\expandafter\ifx\csname citenamefont\endcsname\relax
  \def\citenamefont#1{#1}\fi
\expandafter\ifx\csname url\endcsname\relax
  \def\url#1{\texttt{#1}}\fi
\expandafter\ifx\csname urlprefix\endcsname\relax\def\urlprefix{URL }\fi
\providecommand{\bibinfo}[2]{#2}
\providecommand{\eprint}[2][]{\url{#2}}

\bibitem[{\citenamefont{Datta}(1997)}]{dattabook}
\bibinfo{author}{\bibfnamefont{S.}~\bibnamefont{Datta}},
  \emph{\bibinfo{title}{Electronic Transport in Mesoscopic Systems}}
  (\bibinfo{publisher}{Cambridge University Press}, \bibinfo{year}{1997}).

\bibitem[{\citenamefont{Haug and Jauho}(1996)}]{JauhoBook}
\bibinfo{author}{\bibfnamefont{H.}~\bibnamefont{Haug}} \bibnamefont{and}
  \bibinfo{author}{\bibfnamefont{A.-P.} \bibnamefont{Jauho}},
  \emph{\bibinfo{title}{Quantum Kinetics in Transport and Optics of
  Semiconductors}} (\bibinfo{publisher}{Springer Verlag},
  \bibinfo{year}{1996}).

\bibitem[{\citenamefont{Taylor et~al.}(2001)\citenamefont{Taylor, Guo, and
  Wang}}]{Taylor1}
\bibinfo{author}{\bibfnamefont{J.}~\bibnamefont{Taylor}},
  \bibinfo{author}{\bibfnamefont{H.}~\bibnamefont{Guo}}, \bibnamefont{and}
  \bibinfo{author}{\bibfnamefont{J.}~\bibnamefont{Wang}},
  \bibinfo{journal}{Phys.Rev. B} \textbf{\bibinfo{volume}{63}},
  \bibinfo{pages}{245407} (\bibinfo{year}{2001}).

\bibitem[{\citenamefont{Brandbyge et~al.}(2002)}]{Taylor2}
\bibinfo{author}{\bibfnamefont{M.}~\bibnamefont{Brandbyge}}
  \bibnamefont{et~al.}, \bibinfo{journal}{Phys. Rev. B}
  \textbf{\bibinfo{volume}{65}}, \bibinfo{pages}{165401}
  (\bibinfo{year}{2002}).

\bibitem[{\citenamefont{Jansen et~al.}(1980)\citenamefont{Jansen, van Gelder,
  and Wyder}}]{Jansen1980}
\bibinfo{author}{\bibfnamefont{A.~G.~M.} \bibnamefont{Jansen}},
  \bibinfo{author}{\bibfnamefont{A.~P.} \bibnamefont{van Gelder}},
  \bibnamefont{and} \bibinfo{author}{\bibfnamefont{P.}~\bibnamefont{Wyder}},
  \bibinfo{journal}{J. Phys. C: Solid St. Phys.} \textbf{\bibinfo{volume}{13}},
  \bibinfo{pages}{6073} (\bibinfo{year}{1980}).

\bibitem[{\citenamefont{Ng}(1992)}]{Ng92}
\bibinfo{author}{\bibfnamefont{T.~K.} \bibnamefont{Ng}},
  \bibinfo{journal}{Phys. Rev. Lett.} \textbf{\bibinfo{volume}{68}},
  \bibinfo{pages}{1018} (\bibinfo{year}{1992}).

\bibitem[{\citenamefont{Heinonen and Johnson}(1993)}]{Heinonen93}
\bibinfo{author}{\bibfnamefont{O.}~\bibnamefont{Heinonen}} \bibnamefont{and}
  \bibinfo{author}{\bibfnamefont{M.~D.} \bibnamefont{Johnson}},
  \bibinfo{journal}{Phys. Rev. Lett.} \textbf{\bibinfo{volume}{71}},
  \bibinfo{pages}{1447} (\bibinfo{year}{1993}).

\bibitem[{\citenamefont{Bokes and Godby}(2003)}]{Bokes03}
\bibinfo{author}{\bibfnamefont{P.}~\bibnamefont{Bokes}} \bibnamefont{and}
  \bibinfo{author}{\bibfnamefont{R. W}~\bibnamefont{Godby}},
  \bibinfo{journal}{Phys. Rev. B} \textbf{\bibinfo{volume}{68}},
  \bibinfo{pages}{125414} (\bibinfo{year}{2003}).

\bibitem[{\citenamefont{Bokes et~al.}(2005)\citenamefont{Bokes, Mera, and
  Godby}}]{Bokes05a}
\bibinfo{author}{\bibfnamefont{P.}~\bibnamefont{Bokes}},
  \bibinfo{author}{\bibfnamefont{H.}~\bibnamefont{Mera}}, \bibnamefont{and}
  \bibinfo{author}{\bibfnamefont{R.~W.} \bibnamefont{Godby}},
  \bibinfo{journal}{Phys. Rev. B} \textbf{\bibinfo{volume}{72}},
  \bibinfo{pages}{165425} (\bibinfo{year}{2005}).

\bibitem[{\citenamefont{Todorov et~al.}(2000)\citenamefont{Todorov, Hoekstra,
  and Sutton}}]{Todorov00}
\bibinfo{author}{\bibfnamefont{T.~N.} \bibnamefont{Todorov}},
  \bibinfo{author}{\bibfnamefont{J.}~\bibnamefont{Hoekstra}}, \bibnamefont{and}
  \bibinfo{author}{\bibfnamefont{A.~P.} \bibnamefont{Sutton}},
  \bibinfo{journal}{Phil. Mag. B} \textbf{\bibinfo{volume}{80}},
  \bibinfo{pages}{421} (\bibinfo{year}{2000}).

\bibitem[{\citenamefont{Hershfield}(1993)}]{Hershfield93}
\bibinfo{author}{\bibfnamefont{S.}~\bibnamefont{Hershfield}},
  \bibinfo{journal}{Phys. Rev. Lett.} \textbf{\bibinfo{volume}{70}},
  \bibinfo{pages}{2134} (\bibinfo{year}{1993}).

\bibitem[{\citenamefont{Skudlarski and Vignale}(1993)}]{sk93}
\bibinfo{author}{\bibfnamefont{P.}~\bibnamefont{Skudlarski}} \bibnamefont{and}
  \bibinfo{author}{\bibfnamefont{G.}~\bibnamefont{Vignale}},
  \bibinfo{journal}{Phys. Rev. B} \textbf{\bibinfo{volume}{48}},
  \bibinfo{pages}{8547} (\bibinfo{year}{1993}).

\bibitem[{\citenamefont{Mera et~al.}(2005)\citenamefont{Mera, Bokes, and
  Godby}}]{Mera05}
\bibinfo{author}{\bibfnamefont{H.}~\bibnamefont{Mera}},
  \bibinfo{author}{\bibfnamefont{P.}~\bibnamefont{Bokes}}, \bibnamefont{and}
  \bibinfo{author}{\bibfnamefont{R.~W.} \bibnamefont{Godby}},
  \bibinfo{journal}{Phys. Rev. B} \textbf{\bibinfo{volume}{72}},
  \bibinfo{pages}{085311} (\bibinfo{year}{2005}).

\end{thebibliography}
\end{document}